\documentclass[5p,times,nopreprintline]{elsarticle}

\usepackage{amsmath, amsthm, amssymb, amsfonts, amsbsy, mathrsfs}

\usepackage{xcolor}
\usepackage{calligra,bm}
\usepackage{stmaryrd}
\usepackage{yfonts}
\usepackage{accents}

\usepackage{graphicx}
\graphicspath{{./}}

\usepackage[hidelinks,bookmarks=true]{hyperref} 
\hypersetup{pdfstartview=FitH,pdfhighlight=/O,colorlinks=false}
\usepackage{orcidlink}

\begin{document}

\begin{frontmatter}

\title{\raggedright Aiming for Proxima Centauri b: Gravitational effects on relativistic spacecraft trajectories}

\author[inst1a,inst1b]{Mark C. Baumann\orcidlink{0000-0002-8713-3142}}
\address[inst1a]{Department of Mathematics, St. Edward's University, Austin, TX}
\address[inst1b]{Department of Physics, The University of Texas at Austin, Austin, TX}
\ead{mark@markbaumann.net}
\author[inst2]{Nicky Ishaak\orcidlink{0000-0002-2345-7121}}
\ead{njishaak@outlook.com}
\address[inst2]{Independent Researcher, Austin, TX}
\author[inst3]{Justin C. Feng\orcidlink{0000-0003-2441-5801}}
\ead{feng@fzu.cz}
\address[inst3]{Central European Institute for Cosmology and Fundamental
    Physics, Institute of Physics of the Czech Academy of Sciences, Na 
    Slovance 1999/2, 182 21 Prague 8, Czech Republic}



%
%

%
%
\begin{abstract}
   How important are gravitational and relativistic effects for interstellar travel? We consider this question in the context of proposed laser-propelled
   spacecraft missions to neighboring stellar destinations. In this work, we employ a Julia reimplementation of the \texttt{PoMiN} code, an N-body code modeling relativistic gravitational dynamics in the first post-Minkowskian (PM) approximation to general relativity (valid to linear order in Newton's constant $G$). We compute the gravitational influence of seven different celestial bodies and find that the Sun has the greatest influence on the trajectory of the interstellar spacecraft.  We also study the differences between Newtonian and PM gravity, and find that if mission planners wish to hit Proxima Centauri b with an accuracy of better than about 690,000 kilometers, relativistic effects must be taken into account. To solve for the precise initial data needed to hit an intended target, we develop numerical fine-tuning methods and demonstrate that these methods can (within a given model) be precise to about a femtometer over a travel distance of $\sim4.25$ light years. However, we find that for the spacecraft trajectories we consider, higher order general relativistic effects (beyond the first PM approximation) from the Sun can displace the final position of the spacecraft by tens of kilometers. We also consider the variation in the initial direction of the spacecraft velocity and find that, even with relativistic effects properly taken into account, the miss distances can be dominated by the variation in the initial velocity that could arise from errors during the launch and boost phase of the spacecraft mission.

\end{abstract}


%
%

\end{frontmatter}


%
%

%
%

\section{Introduction}

The Breakthrough Starshot \cite{Lubin:2015,Lubin:2016roadmap,Lubin:2024large,Parkin:2018breakthrough} and NASA Starlight \cite{Kulkarni:2018} projects propose building a laser-propelled spacecraft that can reach relativistic speeds \cite{Marx:1966}. In these proposals, a gram-scale spacecraft is attached to a gram-scale photon sail and is accelerated to $0.2c$ using a ground- or space-based laser array. At that speed, it would take about 20 years to traverse the 4.25 light years to Proxima Centauri, our nearest neighboring star and planetary system. A particularly interesting target is the planet Proxima Centauri b, which is a terrestrial-sized planet in the habitable zone around Proxima Centauri \cite{Brugger:2016,Lovis:2017}.  More recently, similar methods were proposed for a mission to the nearest black hole, which has been estimated to be as close as 20-25 light years away \cite{Bambi:2025kcr,Bambi:2025hjn} (in this article, we will focus on a mission to Proxima Centauri b).

The spacecraft described in these proposals each consist of a microchip with sensors and communicators and no propulsion of its own besides possibly a miniature photon drive. After the initial acceleration, the spacecraft is left to make the $\sim20$-year voyage to Proxima Centauri (or perhaps a longer voyage to a black hole) unaided aside from tiny course corrections. The accurate and precise modeling of the spacecraft's trajectory over such long distances is critical for the success of the mission.

In this article, we consider the gravitational contributions to such spacecraft trajectories. One would expect the trajectory to depend on the gravitational interactions between a number of bodies, including but not limited to: the spacecraft itself, the Earth, the Sun, the Moon, Jupiter, Mars, and the stars of the Alpha Centauri triple star system. Additionally, a model is required that can account for relativistic effects, as the spacecraft is moving at a significant fraction of the speed of light.

$\texttt{PoMiN}$ is an ideal tool for calculating the gravitational effects on the trajectory of such a spacecraft. $\texttt{PoMiN}$ is a relativistic $N$-body gravitational dynamics solver in the first post-Minkowskian approximation to general relativity (GR). Here, ``Minkowskian'' refers to the Minkowski spacetime of special relativity, and post-Minkowskian (PM) refers to results obtained from a perturbative expansion (over the Minkowski spacetime) of the Einstein field equations of GR in orders of Newton's gravitational constant $G$. The first PM approximation therefore corresponds to first-order gravitational corrections to special relativistic dynamics in $G$ \cite{Bjerrum-Bohr:2022blt,Blanchet:2013haa,Detweiler:1996mq} and is accurate up to ultra-relativistic speeds. Such approximations have recently become a topic of intense activity due to their applications in source modeling for gravitational wave observations and recent advances in quantum field theoretical techniques---an overview of recent developments in the field can be found in chapters 13 and 14 of the SageX review on scattering amplitudes \cite{Travaglini:2022uwo,Bjerrum-Bohr:2022blt,Kosower:2022yvp}. 

Ledvinka, Sch{\"a}fer, and Bi{\v c}{\'a}k (hereafter LSB) derived an $N$-body Hamiltonian in the first PM approximation to GR in \cite{PM}. $\texttt{PoMiN}$ computes $N$-body dynamics by numerically solving Hamilton's equations using the PM Hamiltonian of LSB. As the spacecraft is traveling at relativistic speeds and its trajectory remains distant from gravitating objects (compared to their Schwarzschild radii), the PM approximation is appropriate for the accurate modeling of the gravitational dynamics.

$\texttt{PoMiN}$ was originally written in the C language \cite{Feng:2018}, but has recently been ported to the Julia language, which offers greater flexibility in the choice of numerical methods as well as greater compatibility with existing automatic differentiation libraries.  This study provides the first use of the Julia-based $\texttt{PoMiN}$ code, which can be found in \cite{Feng:2018_17246109}.  

In this work, we aim to establish the relativistic and gravitational contributions to the requirements analysis for laser-propelled spacecraft trajectories, and as such, it addresses gravitational dynamics only. Additional forces such as radiation pressure and Lorentz forces have been addressed elsewhere \cite{Hoang2016,Hoang2017,Jackson:2017disp}. 

This article is organized as follows. In section \ref{sec:TargKin}, we discuss the targeting problem for linear trajectories in special relativity. In section \ref{sect:targeting-with-gravity}, we consider the targeting problem in the presence of the PM approximation to GR and describe our numerical methods for solving the targeting problem and how our approximate methods quantitatively differ from the exact predictions of GR. We then discuss the numerical setup in section \ref{sec:Numerical-setup}, and present the results of our simulations in section \ref{sec:simulation-results}. Finally, we present our conclusions in section \ref{sec:conclusions}.


%
%

%
%
%
\section{The Kinematical Targeting Problem}\label{sec:TargKin}
We are interested in solving the targeting problem, which is the problem of finding the initial data that one must supply to a dynamical system in order for the system to pass through a specific point at a given time.

We begin by considering relativistic particles moving in the absence of gravity. Constant velocity motion can be described with the formulas of nonrelativistic kinematics. To see this, we start with the formulas for relativistic kinematics and show that it reduces to the nonrelativistic case.  Consider constant velocity motion parameterized by proper time $\tau$:
\begin{equation} \label{eqA:coordparam}
    X^\mu(\tau) = X^\mu_0 + U^\mu \tau,
\end{equation}
where $X^\mu(\tau)$ are the spacetime coordinates of the particle as a function of its proper (clock) time $\tau$, and $\mu$ is a coordinate index (assuming rectangular coordinates), with $\mu=0$ denoting the time coordinate, and $\mu\in\{1,2,3\}$ denoting spatial coordinates. The quantity $X^\mu_0$ is the initial position, and $U^\mu$ is the (constant) four-velocity $U^\mu = d X^\mu/d\tau$, the components of which may be written:
\begin{equation}
    U^\mu = \left(c \gamma, \gamma \mathbf{v}\right), \qquad \gamma:=\frac{dt}{d\tau}=\frac{1}{\sqrt{1-v^2}},
\end{equation}
where $\mathbf{v}=d\mathbf{x}/dt$ is the velocity vector (with length $v:=\sqrt{\mathbf{v}\cdot\mathbf{v}}$) and $\mathbf{x}(t)$ form the spatial coordinates of the particle. The reader may recognize the quantity $\gamma$ to be the Lorentz factor in special relativity, which is constant if $v$ is constant. The proper time $\tau$ is then linearly related to the coordinate time $t$, and if $\tau=0$ is synchronized to $t=0$, the spatial part of Eq. \eqref{eqA:coordparam} yields the familiar (nonrelativistic) kinematic form:
\begin{equation} \label{eqA:3dparam}
    \mathbf{x}(t) = \mathbf{x}_0 + \mathbf{v} t.
\end{equation}
It follows that in the absence of gravity, the targeting problem can be treated in an entirely nonrelativistic manner.

\begin{figure}[ht]
\includegraphics[width=\columnwidth]{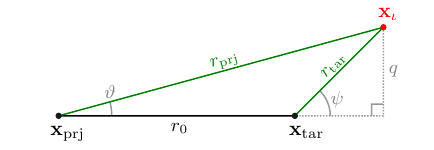}
\caption{Illustration of the geometry of the targeting problem, in the plane formed by the velocity vectors $\mathbf{v}_\mathrm{prj}$ and $\mathbf{v}_\mathrm{tar}$.}
\label{fig:triangles}
\end{figure}

We now consider the problem of hitting a target with a projectile in the case where both the target and projectile are moving at a constant velocity. Let $\mathbf{x}_\mathrm{prj}$ be the initial position of the projectile, and $\mathbf{x}_\mathrm{tar}$ be the initial position of the target. The target is moving with a constant velocity $\mathbf{v}_\mathrm{tar}$, and the magnitude of the projectile's velocity $v_\mathrm{prj}=|\mathbf{v}_\mathrm{prj}|$ is known. The problem is to determine the direction of $\mathbf{v}_\mathrm{prj}$, given $\mathbf{x}_\mathrm{prj}$, $\mathbf{x}_\mathrm{tar}$, $\mathbf{v}_\mathrm{tar}$, and $\mathbf{v}_\mathrm{tar}$. The first observation to make is that if the projectile and target meet, then their trajectories meet at a point $\mathbf{x}_\iota$ at some time $t_\iota$, so that:
\begin{equation} \label{eqA:3dparammeet}
    \mathbf{x}_\iota = \mathbf{x}_\mathrm{prj} + \mathbf{v}_\mathrm{prj} t_\iota = \mathbf{x}_\mathrm{tar} + \mathbf{v}_\mathrm{tar} t_\iota,
\end{equation}
which implies that the differences
$\Delta \mathbf{x}:=\mathbf{x}_\mathrm{prj}-\mathbf{x}_\mathrm{tar}$, and  $\Delta \mathbf{v}:=\mathbf{v}_\mathrm{prj}-\mathbf{v}_\mathrm{tar}$ are proportional:
\begin{equation} \label{eqA:3dparamdiff}
    \Delta \mathbf{x} = \Delta \mathbf{v} t_\iota.
\end{equation}
Moreover, the three points $\mathbf{x}_\mathrm{prj}$, $\mathbf{x}_\mathrm{tar}$, and $\mathbf{x}_\iota$ form a triangle, as illustrated in Figure \ref{fig:triangles}, with sides $r_0:=|\Delta \mathbf{x}|$, $r_\mathrm{prj}:=|\mathbf{x}_\iota-\mathbf{x}_\mathrm{prj}|$, and $r_\mathrm{tar}:=|\mathbf{x}_\iota-\mathbf{x}_\mathrm{tar}|$. It follows that the vector $\mathbf{v}_\mathrm{tar}$ must lie in the same plane as this triangle, differing from $\Delta \mathbf{x}$ by an angle $\vartheta$. Since the projectile and target travel the respective distances $r_\mathrm{prj}$ and $r_\mathrm{tar}$ in the same time $t_\iota$, one can solve for the time of flight and eliminate it to obtain:
\begin{equation} \label{eqA:vrratio}
     \frac{r_\mathrm{tar}}{r_\mathrm{prj}}=\frac{v_\mathrm{tar}}{v_\mathrm{prj}}.
\end{equation}
We now have all the information needed to determine $\vartheta$. From Figure \ref{fig:triangles}, we have:
\begin{equation} \label{eqA:qrel}
     q= r_\mathrm{prj} \sin(\vartheta) = r_\mathrm{tar} \sin(\psi),
\end{equation}
where $\psi$ is the angle formed between the vector $\mathbf{v}_\mathrm{tar}\propto\mathbf{x}_\iota-\mathbf{x}_\mathrm{tar}$ and $-\Delta \mathbf{x}$, which satisfies the dot product relation:
\begin{equation} \label{eqA:cosrel}
     {v}_\mathrm{tar} r_0 \cos(\psi) = - \mathbf{v}_\mathrm{tar} \cdot \Delta \mathbf{x},
\end{equation}
we arrive at the following angle formulas:
\begin{equation}
    \vartheta=\arcsin\left(\frac{v_\mathrm{tar}}{v_\mathrm{prj}}\sin(\psi)\right), \qquad\psi=
     \arccos\left(\frac{- \mathbf{v}_\mathrm{tar} \cdot \Delta \mathbf{x}}{{v}_\mathrm{tar} r_0}\right).
\end{equation}
To construct the vector $\mathbf{v}_\mathrm{prj}$, we define two unit basis vectors $\mathbf{e}_\parallel:=-\Delta \mathbf{x}/r_0$ and $\mathbf{e}_\perp$, the latter being defined as:
\begin{equation}
    \mathbf{e}_\perp := \frac{\mathbf{v}_\mathrm{tar}-(\mathbf{e}_\parallel\cdot\mathbf{v}_\mathrm{tar})\mathbf{e}_\parallel}{v_\mathrm{tar}}.
\end{equation}
These basis vectors span the plane of the triangle in Figure \ref{fig:triangles}. The vector $\mathbf{v}_\mathrm{prj}$ is then given by:
\begin{equation}\label{Eq:FlatSpacev}
    \mathbf{v}_\mathrm{prj} := {v}_\mathrm{prj} \left( \cos(\vartheta) \mathbf{e}_\parallel + \sin(\vartheta) \mathbf{e}_\perp \right).
\end{equation}
This formula can be used to construct the initial velocity for the projectile, given the initial positions of the projectile and the target, and the velocity of the target.

The targeting problem is less straightforward when gravitational interactions are included. Even in the case of Newtonian gravity, the exact problem becomes analytically intractable when the number of gravitating particles exceeds two. Moreover, a velocity of magnitude $v=0.2c$ is in the relativistic regime, so that an accurate modeling of spacecraft trajectories must include relativistic considerations. In the remainder of this article, we quantify such effects.

%
%
%

\section{The Targeting Problem in the Presence of Gravity}
\label{sect:targeting-with-gravity}

\subsection{Newtonian Gravity}
We now consider the targeting problem in the presence of gravity. In many astrophysical settings, Newtonian gravity suffices to describe the motion of celestial bodies. Specifically, Newtonian gravity provides an excellent description of gravitational dynamics for bodies that are separated by distances much larger than the Schwarzschild radii of the bodies, and for velocities much slower than the speed of light. Of course, the spacecraft is traveling at a significant fraction of the speed of light, and the relevant celestial bodies (such as the Sun and Proxima Centauri) are traveling at relative velocities on the order of $10^{-5} c$, which may be relevant for travel distance of a few light years (on the order of $10^{16}~\mathrm{m}$). Despite these considerations, it is prudent to first model gravitational interactions with Newtonian gravity for comparison.

We employ the Hamiltonian formalism for Newtonian gravity (for a pedagogical review of the Hamiltonian formalism, we refer the reader to the treatments in \cite{Thornton:2004,GoldsteinCM}). In particular, we consider the N-body Hamiltonian for a system of $N$ particles with masses $m_a$:
\begin{equation} \label{NewtHamiltonian}
  H_\mathrm{Newt} (\textbf{x}_a,\vec{p}_a) 
  =
  \sum_{a= 1}^{N} \frac{\vec{p}_a^2}{2 m_a} - \sum_{a,b\neq a}^{N}\frac{G m_a  m_b}{r_{ab}},
\end{equation}
where $\vec{p}_a = m_a \textbf{v}_a$ is the momentum of particle $a$ (with $\textbf{v}_a$ being the velocity of the particle), and $r_{ab}:=|\textbf{x}_a - \textbf{x}_b|$ is the separation distance between particle $a$ and $b$. The Hamiltonian evaluates to the total (kinetic plus potential) energy of the gravitating system, and the motion of each particle in the system is given by Hamilton's equations:
\begin{equation} \label{HamEqs}
\begin{aligned} 
\frac{d\textbf{x}_a}{dt} = \frac{\partial H}{\partial \textbf{p}_a}  ,
\qquad
\frac{d\textbf{p}_a}{dt} = -\frac{\partial H}{\partial \textbf{x}_a} ,
\end{aligned}
\end{equation}
which in general form a system of first order, nonlinear and coupled ordinary differential equations (ODEs).

\subsection{Post-Minkowskian Hamiltonian}
If the system of bodies contains a subset of bodies that move at a nonnegligible fraction of the speed of light, one would prefer a model of gravity that properly accounts for relativistic motion. The leading theory of gravity is general relativity (GR) \cite{MTW,Carroll,Wald}, which is well-established on astrophysical scales \cite{Will:2014kxa,Yunes:2013dva}. However, the nonlinearity of the Einstein field equations render an exact analysis intractable. An approximation is necessary. Since we are considering relativistic velocities and the field of many (weakly) gravitating objects, it is sufficient for our purposes to consider the first post-Minkowskian (PM) approximation\footnote{Since we are considering relativistic motion, we prefer the PM approximation to the post-Newtonian approximation.} to
GR, that is, one in which gravitational effects beyond leading order in Newton's gravitational constant $G$ are discarded \cite{PoissonWill,Detweiler:1996mq}. 
In LSB \cite{PM}, the following $N$-body PM Hamiltonian was
obtained within the framework of the Arnowitt-Deser-Misner (ADM) formulation of GR \cite{ADM62,ADM62b} and the linearized Einstein equations:
\begin{equation} \label{LSBHamiltonian}
  H (\textbf{x}_a,\textbf{p}_a) 
  =
  H_1+H_2+H_3,
\end{equation}
where (for convenience) we have decomposed the Hamiltonian into the following parts:
\begin{equation} \label{LSBHamiltonianH1H2}
\begin{aligned} 
  H_1 = \sum_{a= 1}^{N}{E}_{a},
  \quad
  H_2 = - \frac{1}{2}G\sum_{a,b\neq a}^{N}\frac{{E}_{a}{E}_{b}}{r_{ab}}\left(1+ \frac{\textbf{p}^{2}_{a}}{{E}^{2}_{a}}+ \frac{\textbf{p}^{2}_{b}}{{E}^{2}_{b}}\right),
\end{aligned}
\end{equation}
\begin{equation} \label{LSBHamiltonianH3}
\begin{aligned} 
  H_3 
  =&\,
  \frac{1}{4}G\sum_{a,b\neq a}^{N} \frac{1}{r_{ab}}\biggl\{\left(7\Xi_{ab}+ (\mathbf{p}_{a}\cdot\mathbf{n}_{ab})\Theta_{ba}\right) 
  -
  \frac{\left({E}_a{E}_b\right)^{-1}}{\left(y_{ba}+1\right)^{2}y_{ba}}\biggl[ \\
  &
  \left(\mathbf{p}^{2}_{a}\mathbf{p}^{2}_{b} - 3\mathbf{p}^{2}_a\Theta_{ba}^{2}+ \Theta_{ab}^{2}\Theta_{ba}^{2}+ 8\Theta_{ab}\Theta_{ba}\Xi_{ab}-3\Theta_{ab}^{2}\mathbf{p}^{2}_{b}\right)y_{ba}\\
  &
  +2\left(2\Xi_{ab}^{2}\Theta_{ba}^{2}+ 2\Theta_{ab}\Theta_{ba}\Xi_{ab}\mathbf{p}^{2}_{b}+ \Theta_{ab}^{2}\mathbf{p}^{4}_{b} - \Xi_{ab}^{2}\mathbf{p}^{2}_{b}\right)\frac{1}{{E}^{2}_{b}}  \\
  &
  + 2\left(\Xi_{ab}^{2}-\mathbf{p}^{2}_{a}\Theta_{ba}^{2}+ \Theta_{ab}^{2}\Theta_{ba}^{2}- 2\Theta_{ab}\Theta_{ba}\Xi_{ab}- \Theta_{ab}^{2}\mathbf{p}^{2}_{b}\right)\biggr]\biggr\} .
\end{aligned}
\end{equation}
where $\textbf{x}_a$ and $\textbf{p}_a$ are the respective coordinates and momenta of particle $a$, $G$ is Newton's constant, the speed of light is unity ($c=1$), and the following quantities are defined:
\begin{equation} \label{DistAngleDot}
\begin{aligned} 
&r_{ab}:=|\textbf{x}_a - \textbf{x}_b|,\qquad
&\textbf{n}_{ab} :=r_{ab}^{-1}\,(\textbf{x}_a - \textbf{x}_b),\\
&\Theta_{ab} := \mathbf{p}_{a}\cdot\mathbf{n}_{ba},
&\Xi_{ab} := \mathbf{p}_{a}\cdot\mathbf{p}_{b},
\end{aligned}
\end{equation}
\begin{equation} \label{SC2-LSBHamiltonianDefns}
\begin{aligned} 
E_a :=\sqrt{m_a^2+\textbf{p}_a^2},\qquad
y_{ba}:=E_b^{-1}\,
    \sqrt{m^{2}_{b}+\left(\mathbf{n}_{ba} \cdot \mathbf{p}_{b}\right)^{2}}.
\end{aligned}
\end{equation}
The Hamiltonian (hereafter referred to as the LSB Hamiltonian) was derived from the linearized Einstein equations, treating the acceleration of the sources as higher order effects to be neglected; in effect, it contains all corrections linear in $G$. The momenta are relativistic, that is, they are related to the velocities $\textbf{v}_a$ according to $\textbf{p}_a=\gamma_a m_a \textbf{v}_a$, where the Lorentz factor is defined:
\begin{equation} \label{LorentzFactor}
\gamma_a := \frac{1}{\sqrt{1-\textbf{v}_a^2}}.
\end{equation}
A momentum exchange formula for the scattering between two particles was also derived in \cite{PM}, assuming the scattering process results in slight deviations from the straight line motion of the particles:
\begin{equation}
\begin{aligned} \label{delta_p}
\Delta \textbf{p} = -2\frac{{\textbf{b}}}{{\textbf{b}}^2} \frac{G}{p}
\frac{{E}_1^2 {E}_2^2}{{E}_1 + {E}_2 }
\biggl[&
1+\left(\frac{1}{{E}_1^2}+\frac{1}{{E}_2^2}+\frac{4}{{E}_1 {E}_2} \right){p}^2 + \frac{{p}^4}{{E}_1^2 {E}_2^2 }
\biggr]~.
\end{aligned}
\end{equation}
In \cite{Doss:2016convergence}, the well-known general relativistic prediction for the light deflection angle was recovered for the LSB Hamiltonian. These results are used for the validation of the code and methodology in the next subsection.

\subsection{Numerical Implementation}
$\texttt{PoMiN}$ was originally implemented in C \cite{Feng:2018izi}, and has recently been reimplemented in the Julia language \cite{Feng:2018_17246109}. We have reimplemented $\texttt{PoMiN}$ in the Julia language for several reasons: the Julia library $\texttt{OrdinaryDiffEq.jl}$ contains state-of-the-art integrators (including symplectic integrators) with built-in interpolation, code implemented in the Julia language is differentiable, and the Julia language offers more flexibility in floating point types.

In the Julia reimplementation, Hamilton's equations are constructed by way of dual number automatic differentiation (AD) using the $\texttt{ForwardDiff.jl}$ library \cite{ForwardDiff}, which implements the so called ``forward mode'' method of AD. This method is based on a dual number type \cite{Burton:2007numerical,Neidinger:2010,Baydin:2018} of the form:\footnote{Colloquially, one may regard dual numbers as being similar to complex numbers, except with the imaginary number $i$ replaced with a one-dimensional Grassmann number $\varepsilon$.}
\begin{equation}
    \begin{aligned}
    z = x + \varepsilon y,
    \end{aligned}
\end{equation}
where $x$ and $y$ are real numbers, and $\varepsilon$ is a one-dimensional Grassman number defined by the property $\varepsilon^2=0$. Given this property, one can show (via a Taylor expansion) that any analytic function $f(x)$ satisfies the following expression:
\begin{equation}
    \begin{aligned}
    f(z) = f(x) + \varepsilon y \frac{df}{dx}.
    \end{aligned}
\end{equation}
One may then implement dual numbers in a manner similar to that for complex numbers, so that machine precision derivatives for functions can be computed with a computational efficiency similar to that of the evaluation of analytic functions on complex numbers.

For the integration of Hamilton's equations, we have included both a reimplementation of the Runge-Kutta scheme from the C code (with interaction-dependent adaptive timestepping), as well as our default option to employ a method from the $\texttt{OrdinaryDiffEq.jl}$ library \cite{DifferentialEquations.jl-2017}. The results in this work were obtained using the \texttt{Vern7} and the \texttt{Vern9} methods \cite{Verner:2010numerically} (Verner's efficient 7/6 Runge-Kutta method and Verner's efficient 9/8 Runge-Kutta method, using 7th order and 9th order interpolants, respectively), and the available testing suite was used to validate the routines in $\texttt{OrdinaryDiffEq.jl}$.

\begin{figure}[ht]
\includegraphics[width=\columnwidth]{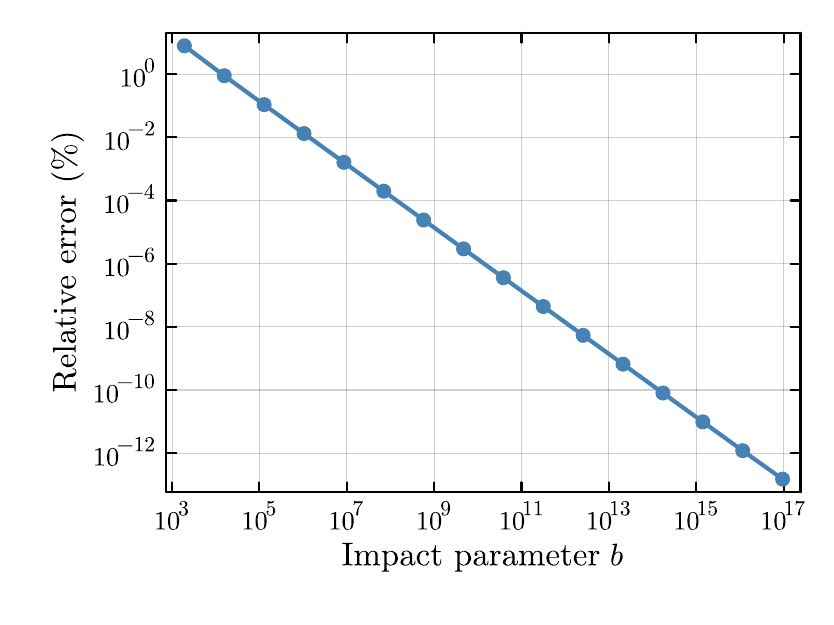}
\caption{The relative error $|\Delta \textbf{p}_\mathrm{num}-\Delta \textbf{p}_\mathrm{an}|/|\Delta \textbf{p}_\mathrm{an}|$ between the numerical momentum exchange $\Delta \textbf{p}_\mathrm{num}$ and the analytical prediction $\Delta \textbf{p}_\mathrm{an}$ of Eq. \eqref{delta_p} versus the impact parameter $b$. The scaling behavior indicates that the numerical result converges to the analytical prediction at a rate of $\sim1/b$, as one might expect from Eq. \eqref{delta_p}.}
\label{fig:MXerror}
\end{figure}

Comparison tests with the momentum exchange formula \eqref{delta_p} were performed, with the Julia integrators exceeding the performance of the hand-coded Runge-Kutta integrator in the original C implementation \cite{Feng:2018izi}. In particular, the power-law scaling for the numerical errors can be extended to arbitrarily large values of the impact parameters by decreasing the tolerances and by using extended precision floating point types (such as those supplied in the $\texttt{DoubleFloats.jl}$ library \cite{DoubleFloats2018}). A representative plot is given in Figure \ref{fig:MXerror}.

Our simulations divide the bodies into two subsets. The first subset consists of $N$ fully interacting bodies governed by either the Newtonian Hamiltonian $H_\mathrm{Newt}$ in Eq. \eqref{NewtHamiltonian} or the LSB Hamiltonian $H$ in Eqs. (\ref{LSBHamiltonian}--\ref{LSBHamiltonianH3}). The second subset consists of test particles. The test particles are each governed by a $N+1$ body Hamiltonian that includes the contribution of the test particle. However, the test particles themselves do not contribute to the Hamiltonian of the first subset, nor to the Hamiltonians of other distinct test particles.

\subsection{Numerical Fine Tuning of Initial Data}\label{sec:FineTuning}

As we have determined in section \ref{sec:TargKin}, the targeting problem in the absence of gravity can be treated nonrelativistically, even if the particles are moving with relativistic velocities. When gravitational effects are included, one might expect general relativistic effects to become important for fast moving particles---we will later justify this expectation numerically using the LSB Hamiltonian in Eqs. (\ref{LSBHamiltonian}--\ref{LSBHamiltonianH3}). Due to the complexity of the Hamiltonian and the nonlinearity of the resulting set of Hamilton's equations Eq. \eqref{HamEqs}, the problem of obtaining an analytical solution to the targeting problem is intractable. However, the Julia refactoring of \texttt{PoMiN} yields a new strategy for solving the targeting problem to machine precision.

In \cite{Feng:2022imj}, it was demonstrated that existing Julia libraries (the library \texttt{OrdinaryDiffEq.jl}, in particular \cite{DifferentialEquations.jl-2017}) for numerically solving ODEs are sufficiently compatible with AD to permit the differentiation of numerical ODE solutions to Hamilton's equations with respect to initial data. One can use this with generalizations of Newton-Raphson root finding methods to solve the targeting problem. In principle, one can construct a (vector-valued) ``miss'' function $\Delta \textbf{x}_{miss}(\textbf{v}_a)$ which computes the miss distance between the the target point and the endpoint of the numerical solution for Eq. \eqref{HamEqs}, given some initial velocity $\textbf{v}_a$. The targeting problem is then reduced to that of a root finding problem; we find the values of the initial velocity $\textbf{v}_a$ such that the function $\Delta \textbf{x}_{miss}(\textbf{v}_a)$ returns a zero vector, that is, we seek solutions to:
\begin{equation}
    \Delta \textbf{x}_{miss}(\textbf{v}_a)=0.
\end{equation}
Now given a vector-valued function $\textbf{f}(\textbf{v})$ one can employ one of many root-finding algorithms; for our purposes, it is sufficient to employ a quasi-Newton Broyden method \cite{Broyden:1965,Press:2007NR}, which updates the system according to:
\begin{equation} \label{Deltavf}
  \begin{aligned}
  \Delta \textbf{v}_{{\rm i}+1} &= \textbf{v}_{{\rm i}+1} - \textbf{v}_{{\rm i}}= {\bf J}^{-1}_{{\rm i}} \textbf{f}_{\rm i}(\textbf{v}_{{\rm i}}),
  \end{aligned}
\end{equation}
where the inverse Jacobian matrix ${\bf J}^{-1}$ is initially computed using AD, and is updated in subsequent steps according to the Sherman-Morrison formula:
\begin{equation} \label{ShermanMorrison}
  \begin{aligned}
    {\bf J}^{-1}_{{\rm i}+1} 
    &= {\bf J}^{-1}_{\rm i}
                +
                \frac{\Delta \textbf{v}^{T}_{\rm i} 
                      - {\bf J}^{-1}_{\rm i} \, \Delta \textbf{f}_{\rm i} 
                     }
                     {\Delta \textbf{v}^{T}_{\rm i} \, {\bf J}^{-1}_{\rm i} \, 
                      \Delta \textbf{f}_{\rm i} 
                     }
                \Delta \textbf{v}^{T}_{\rm i} \, {\bf J}^{-1}_{\rm i}\\
  \Delta \textbf{f}_{\rm i} &= \textbf{f}_{\rm i}(\textbf{v}_{\rm i}) - \textbf{f}_{\rm i}(\textbf{v}_{{\rm i}-1}) .
  \end{aligned}
\end{equation}
Applied to the function $\Delta \textbf{x}_{miss}(\textbf{v}_a)$, one may use the expression for $\textbf{v}_{\textrm{prj}}$ in Eq. \eqref{Eq:FlatSpacev} to obtain an initial value for $\textbf{v}_a$. One may compute the Jacobian $\textbf{J}$ for $\Delta \textbf{x}_{miss}(\textbf{v}_a)$ at the initial value for $\textbf{v}_a$, which supplies the initial conditions for the Broyden method.  This method may be used to solve the targeting problem in the gravitational case.  Combined with high-precision floating point types, this allows us to obtain fine-tuned initial velocities to a precision many orders of magnitude beyond the regime of applicability for the LSB Hamiltonian in Eqs. (\ref{LSBHamiltonian}--\ref{LSBHamiltonianH3}).

\subsection{Scattering in General Relativity}

Even if we can fine-tune the initial data so that the spacecraft hits the target with machine precision accuracy, one might still expect some inaccuracies to arise due to the fact that we are working in a first PM approximation to GR. To estimate the size of these inaccuracies, we consider the gravitational scattering of a particle from a massive, nonrotating, spherically symmetric object. 

\begin{figure}[ht]
\includegraphics[width=\columnwidth]{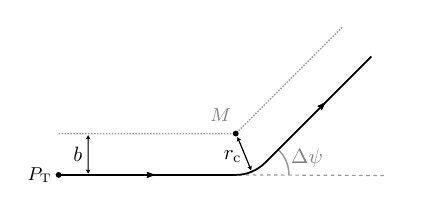}
\caption{Illustration of the gravitational scattering of an incident test particle $P_\mathrm{T}$ from a heavy object of mass $M$. The quantity $b$ is the impact parameter, $r_\mathrm{c}$ is the distance of closest approach, and $\Delta\psi$ is the deflection angle.}
\label{fig:scattering}
\end{figure}

In the scattering problem, we consider an incident test particle $P_\mathrm{T}$ traveling toward a heavy, stationary object of mass $M$ on a trajectory that is (transversely) offset by a distance $b$, commonly known as the impact parameter. In gravitational scattering problems, we typically assume that the particle $P_\mathrm{T}$ is initially placed at a distance far from the center of mass for the stationary object, and ends up at a similarly large distance at late times. As such, we are interested in the angle of deflection $\Delta\psi$ from a straight line trajectory. An illustration of the gravitational scattering problem is provided in Figure \ref{fig:scattering}.

The motion of test particles under the influence of gravity from a nonrotating object of mass $M$ is described by geodesics in the Schwarzschild spacetime, the geometry of which is described by the following line element \cite{MTW,Carroll,Wald}:
\begin{equation}\label{Eq:SchwarzschildLineElement}
    ds^2 = - \left(1-\frac{2GM}{c^2 r}\right) c^2 dt^2 + \frac{dr^2}{1-2GM/(c^2 r)} + r^2 (d\theta^2+\sin^2\theta~d\phi^2),
\end{equation}
written in spherical polar coordinates $x^\mu=(t,r,\theta,\phi)$. Without loss of generality, we may consider motion restricted to the equatorial plane $\theta=\pi/2$. To find geodesics in a given spacetime, one may solve the geodesic equation, but for scattering problems in the Schwarzschild spacetime, this is not necessary, as the invariance of the line element under time translations and rotations can be exploited to obtain the relevant features of the scattering trajectory without directly solving the geodesic equation.

The translational and rotational symmetries yield conserved quantities, the respective specific (per unit mass) energy $e$ and specific angular momentum $l$. The Schwarzschild spacetime is asymptotically flat, so that at large $r$, the line element \eqref{Eq:SchwarzschildLineElement} is equivalent to that of the flat Minkowski spacetime of special relativity. Since the particle is initially placed at large $r$, the energy and angular momentum can then be defined in terms of their special relativistic counterparts constructed from initial data, that is:
\begin{equation}\label{Eq:SpecEnergyAngMom}
    e = \gamma c = c/\sqrt{1-(v/c)^2},
    \qquad
    l = b \gamma v = b \sqrt{e^2-c^2}.
\end{equation}
In terms of these quantities, one can obtain the following equations for the $r$ and $\phi$ coordinates of the geodesic as a function of proper time $\tau$ (cf. Eqs. 6.3.13 and 6.3.14 of \cite{Wald}):
\begin{equation}\label{Eq:GeoEqs}
\begin{aligned}
    \frac{dr}{d\tau}
    &= \sqrt{e^2-\left(1-\frac{2GM}{c^2 r}\right)\left(c^2+\frac{l^2}{r^2}\right)},
    \qquad
    \frac{d\phi}{d\tau}
    = \frac{l}{r^2}.
\end{aligned}
\end{equation}
The radius of closest approach $r_\mathrm{c}$ can be obtained by setting ${dr}/{d\tau}=0$, which establishes the following relation between $l$ and $e$:
\begin{equation}\label{Eq:ApparentImpactParam}
    B:=\frac{l}{e}
      = r_\mathrm{c}
        \sqrt{\frac{r_\mathrm{c}}{r_\mathrm{c}-2GM/c^2}-\frac{c^2}{e^2}},
\end{equation}
where the ratio $B=l/e$ is the apparent impact parameter, which coincides with the impact parameter $b$ in the limit $e \gg c$. One can combine the derivatives in Eqs. \eqref{Eq:GeoEqs} to obtain:
\begin{equation}\label{Eq:AngleODEScattering}
    \frac{dr}{d\phi} 
    = \frac{r}{l}\sqrt{e^2 r^2-(l^2+r^2 c^2)\left(1-\frac{2GM}{c^2 r}\right)},
\end{equation}
yielding the integral:
\begin{equation}\label{Eq:AngleIntegral}
    \begin{aligned}
    \Delta\phi
    &=2\int_{r_\mathrm{c}}^\infty
       \frac{B e ~ dr}{\sqrt{e^2 r(r^3-B^2(r-2 G M/c^2))-c^2 r^3 (r-2G M/c^2)}}.
    \end{aligned}
\end{equation}
The integration limits correspond to a particle that starts at infinity (or infinite radius $r$), reaches a point of closest approach $r_\mathrm{c}$, and goes back out to infinity. One can perform a Taylor expansion of the integrand in powers of $G$ (corresponding to a weak field expansion) and integrate term by term to obtain:
\begin{equation}\label{Eq:IntegralExpand}
    \begin{aligned}
    \Delta\phi
    &=\pi + \Delta\psi,
    \end{aligned}
\end{equation}
with $\Delta\psi$ being the deflection angle:
\begin{equation}\label{Eq:DeflAngle}
    \begin{aligned}
    \Delta\psi
    &=\frac{G M}{r_\mathrm{c} c^2}\frac{c^2+v^2}{v^2}+\frac{G^2 M^2}{r_\mathrm{c}^2 c^4} \frac{\left(3 \pi  v^2 \left(4 c^2+v^2\right)-8 c^2 \left(c^2+v^2\right)\right)}{8 v^4}\\
    &\quad+\mathcal{O}(G^3),
    \end{aligned}
\end{equation}
where we have used the expression $e=1/\sqrt{1-(v/c)^2}$, with $v$ being the velocity of the test particle at infinity. When $G=0$ (gravitational interactions turned off) the deflection angle $\Delta\psi$ angle  $\Delta\phi$ is $\pi$ radians, which corresponds to a $180^\circ$ change in the position of the particle, as one might expect for a straight line trajectory.

It is convenient to obtain an expression for the radius of closest approach $r_\mathrm{c}$ in terms of the impact parameter $b$. One can solve Eq. \eqref{Eq:SpecEnergyAngMom} for $b$, and upon performing the replacement $l=Be$ and setting $e=\gamma c=c/\sqrt{1-(v/c)^2}$, one may perform a Taylor expansion in $G$ of the resulting expression.
The Taylor expansion of $b$ can then be recursively inverted to obtain the following expression for the radius of closest approach $r_\mathrm{c}$ as a function of the impact parameter:
\begin{equation}\label{Eq:rcexpandion}
    r_\mathrm{c} = b - \frac{G M}{v^2} + \frac{G^2 M^2 \left(c^2-4 v^2\right)}{2 b c^2 v^4}
    +\mathcal{O}(G^3) .
\end{equation}
One can then replace occurrences of $r_\mathrm{c}$ with the above to obtain expressions that depend explicitly on the initial data consisting of the impact parameter $b$ and speed $v$.

For a scattering trajectory of a finite length $L$, one can use the deflection angle to estimate the transverse displacement $\Delta x_\perp$ of the gravitationally deflected trajectory from a straight line path according to the formula (which makes use of the small angle approximation $\sin(\Delta\psi)\approx\Delta\psi$):
\begin{equation}\label{Eq:TransverseDispl0}
    \begin{aligned}
    \Delta x_\perp \approx L \Delta\psi = \Delta x_{\perp(G)} + \Delta x_{\perp(G^2)} + \mathcal{O}(G^3),
    \end{aligned}
\end{equation}
where explicitly,
\begin{equation}\label{Eq:TransverseDispl}
    \begin{aligned}
    \Delta x_{\perp(G)} 
    &:= \frac{2 G M \left(c^2+v^2\right) L}{b c^2 v^2},
    \quad
    \Delta x_{\perp(G^2)} 
    = 
    \frac{3 \pi G^2 M^2 \left(4 c^2+v^2\right) L}{4 b^2 c^4 v^4}.
    \end{aligned}
\end{equation}
The expressions for $\Delta x_{\perp(G)}$ and $\Delta x_{\perp(G^2)}$ can be used to estimate the relative order of magnitude contributions of the respective first and second order PM corrections to the displacement (the second order PM corrections being defined as those proportional to $G^2$). In this way, we can account for the difference between the first PM approximation that we employ in our work and higher order effects in GR.

It is useful to generalize our results to the case where the central mass is moving with a constant velocity. When the incident test particle $P_\mathrm{T}$ is far from the central mass, special relativity provides a good approximation to the motion of the test particle and the central mass, so that the kinematical considerations of section \ref{sec:TargKin} apply. If both the test particle $P_\mathrm{T}$ and the central mass are assumed to move on straight lines, and the central mass is moving at velocities much smaller than the speed of light, then the impact parameter corresponds to the point of closest approach. 

We therefore derive an expression for the point of closest approach for two nonintersecting straight line trajectories, $\textbf{x}_{\mathrm{a}}(t)=\textbf{x}_{\mathrm{a},0}+\textbf{v}_{\mathrm{a}} t$ and $\textbf{x}_{\mathrm{b}}(t)=\textbf{x}_{\mathrm{b},0}+\textbf{v}_{\mathrm{b}} t$, where $\textbf{x}_{\mathrm{a},0}=\textbf{x}_{\mathrm{a}}(0)$ and $\textbf{x}_{\mathrm{b},0}=\textbf{x}_{\mathrm{a}}(0)$ are the initial positions and $\textbf{v}_{\mathrm{a}}$ and $\textbf{v}_{\mathrm{b}}$ are velocities. One may define the difference:
\begin{equation}\label{Eq:TwoPathDiff}
    \begin{aligned}
    \Delta \textbf{x}(t)
    :=
    \textbf{x}_{\mathrm{b}}(t)-\textbf{x}_{\mathrm{a}}(t)
    = \Delta \textbf{x}_{0} + \Delta\textbf{v}_{0} t,
    \end{aligned}
\end{equation}
where $\Delta \textbf{x}_{0}:=\textbf{x}_{\mathrm{b},0}-\textbf{x}_{\mathrm{a},0}$ and $ \Delta\textbf{v}_{0}:=\textbf{v}_{\mathrm{b}}-\textbf{v}_{\mathrm{a}}$. The point of closest approach is given by the minimum value of $|\Delta \textbf{x}(t)|^2$, which is given by the time $t_{\textbf{cl}}$ satisfying the following:
\begin{equation}
    \begin{aligned}
    \left.\frac{1}{2}\frac{d |\Delta \textbf{x}(t)|^2}{dt}\right|_{t=t_{\textbf{cl}}}
    =
    \Delta \textbf{x}(t_{\textbf{cl}}) \cdot \Delta\textbf{v}_{0} 
    =
    \Delta \textbf{x}_{0} \cdot \Delta\textbf{v}_{0} 
    + |\Delta\textbf{v}_{0}|^2 t_{\textbf{cl}}
    = 
    0,
    \end{aligned}
\end{equation}
and after solving for $t_{\textbf{cl}}$ and substituting into Eq. \eqref{Eq:TwoPathDiff}, one may obtain the following expression for the distance of closest approach:
\begin{equation} \label{Eq:ClosestApproachFlat}
    \begin{aligned}
    |\Delta \textbf{x}(t_{\textbf{cl}})|
    &= \sqrt{|\Delta \textbf{x}_{0}|^2 - \frac{|\Delta \textbf{x}_{0}\cdot\Delta\textbf{v}_{0}|^2}{|\Delta\textbf{v}_{0}|^2}}.
    \end{aligned}
\end{equation}
Note that this formula depends only on the initial data for the straight line trajectories. If the central mass is moving much slower than the speed of light, then one may estimate the velocity with $v\sim|\Delta \mathbf{v}|$, and the impact parameter with $b\sim|\Delta \textbf{x}(t_{\textbf{cl}})|$. 

%
%

%
%

%
%
%

\section{Numerical Setup}\label{sec:Numerical-setup}

To evaluate the performance and accuracy of these methods for the targeting problem (within the framework of $\texttt{PoMiN}$), we perform several sets of calculations involving a set of gravitating bodies, corresponding to selected celestial objects relevant for modeling gravitational perturbations for a mission to Proxima Centauri b. 

\subsection{The Target: Proxima Centauri b}

An appealing candidate target for a first interstellar mission is Proxima Centauri b, the terrestrial-sized planet in the habitable zone around Proxima Centauri.  The orbit of Proxima Centauri b has a semimajor axis of about 0.05 AU.  In an effort to hit a target like Proxima Centauri b, we investigate the accuracy with which one must launch a relativistic gram-scale spacecraft to hit a point that is located a distance of 0.05 AU from Proxima Centauri.


\subsection{Initial Data}\label{sec:InitialData}


$\texttt{PoMiN}$ is used to calculate the gravitational interactions among up to eight bodies: the relativistic spacecraft, the Earth, the Sun, the Moon, Jupiter, Mars, and the stars of the Alpha Centauri system: Proxima Centauri and the barycenter of Alpha Centauri A and Alpha Centauri B (treated as a single particle).  

We treat the spacecraft itself as a test particle, meaning it moves under the influence of gravity but its own gravitational field is ignored; this is warranted considering the spacecraft has a mass of just 2 grams: a 1 gram chip plus a 1 gram laser sail.  In practice, we vary the mass of the spacecraft by several orders of magnitude and find that the mass of the spacecraft itself has a negligible effect.

\subsection{$N$-Body Initial Conditions}
\label{NbodyInitCond}

The masses and positions and velocities of Proxima Centauri and the Alpha Centauri A+B barycenter are given in Tables 1 and B.1 of \cite{Kervella:2017}.  We use an ICRS Cartesian frame with epoch J2000.


We use the positions and velocities of solar system bodies at time J2000, also in the ICRS frame.  We obtain these using ephemerides computed by astropy \cite{astropy1,astropy2,astropy3}.

As detailed in \cite{Parkin}, a potential method for boosting the spacecraft is to use an Earth-based laser to propel the spacecraft for about 9 minutes.  After the acceleration phase, the spacecraft's velocity will be $0.2c$ and its distance from Earth about $0.1~\mathrm{AU}$.  Since our study focuses on the post-propulsion phase of the spacecraft's journey, the initial position of the spacecraft was therefore taken to be $0.1~\mathrm{AU}$ from Earth in the direction of Proxima Centauri.

\subsection{Randomizing Initial Velocities}
\label{sect:random_velocity}


Determining the initial spacecraft velocity is a major goal of this study, and the methods for obtaining the velocity are described in detail in section \ref{sect:targeting-with-gravity}, the results of which are given in section \ref{sect:fine-tuning} and Table \ref{table:FTIDComponents}.  Once we have an initial velocity that will successfully hit Proxima Centauri b, random perturbations in the initial velocity are introduced and we compute the magnitude of miss distances that arise as a result of errors during the initial launch and boost of the spacecraft.

Random velocities are obtained by first specifying a maximum angular deviation $\theta$.  This is the angle of deviation from the baseline velocity vector derived from the fine-tuning process given in section \ref{sect:fine-tuning}.

Let $\vec{x}_s$ be the initial position of the spacecraft and $\vec{x}_t$ be the final position (at time $t_{cl}$) of the target Proxima Centauri b.  Given $\theta$ and the distance $d = |\vec{x}_t - \vec{x}_s|$ to the target, a target disk of radius $r$ is constructed around the target where $r$ is given by the small angle formula $\tan(\theta) = r / d \approx \theta$.  The target disk is oriented so that it is normal to the vector $\vec{x}_t - \vec{x}_s$.  A point is sampled uniformly from the target disk using a Monte Carlo method in which points are sampled uniformly from a square $(-r, r) \times (-r,r)$ and any point lying outside the disk is discarded.  This point provides the new perturbed target $\vec{x}_{pt}$.  The random initial velocity has the same direction as $\vec{x}_{pt} - \vec{x}_s$ and is rescaled to have the same length as the fine-tuned velocity given in Table \ref{table:FTIDComponents}.




\section{Simulation Results}\label{sec:simulation-results}

Here, we describe the results of three sets of simulations employing the \texttt{PoMiN} code which inform our evaluation of gravitational effects on the trajectory of a relativistic spacecraft traveling toward Proxima Centauri b.

\subsection{Two-body Calculations}
The first set is a series of two-body simulations, each consisting of a massive gravitating body and the spacecraft, with the latter treated as a test particle. The intent is to determine the magnitude of gravitational and relativistic contributions of each massive body to the final position of the spacecraft.

The spacecraft is given an initial velocity that was computed for flat space (i.e., in the absence of gravity) as described in section \ref{sec:TargKin} and we calculate its corresponding miss distance (i.e., the distance of its closest approach to Proxima Centauri b) via simulation.  This ``naive" choice for the initial velocity and the resulting miss distance provide a means of quantifying the gravitational influence of each body individually.

This calculation is done twice, once using the (first) PM approximation to GR (\texttt{PoMiN}), and once using the Newtonian model of gravity.  The resulting miss distances are listed in the first two columns of Table \ref{table:OnePartMisses} and are given in AU.  The rows are ordered from greatest to least effect (in terms of magnitude of miss distance).

The Sun has the greatest effect on the trajectory of the spacecraft, with its influence amounting to a miss distance of about $0.139~\mathrm{AU}$ in the PM gravity calculation.

Also in Table \ref{table:OnePartMisses}, theoretical estimates of miss distances are given for the purpose of comparison.  The theoretical estimates are rough upper limits obtained from the scattering problem in the Schwarzschild spacetime.  With the exception of Proxima Centauri, the estimates in the $\mathcal{O}(G)$ column are obtained from $\Delta x_{\perp(G)}$ in Eq. \eqref{Eq:TransverseDispl}, and the estimates in the $\mathcal{O}(G^2)$ column are obtained from $\Delta x_{\perp(G^2)}$. In the case of Proxima Centauri, the spacecraft is near the radius of closest approach $r_\mathrm{c}$, so that the $\mathcal{O}(G)$ and $\mathcal{O}(G^2)$ estimates for Proxima Centauri are obtained from the magnitude of the corresponding terms in Eq. \eqref{Eq:rcexpandion}. 

The theoretical $\mathcal{O}(G)$ estimates are comparable in order of magnitude to the computed miss distances in the first column of Table \ref{table:OnePartMisses} (as one might expect for the PM approximation), but they differ significantly for the case of the Earth and to a lesser degree for Alpha Centauri A+B. This is partly because the trajectory of the spacecraft does not pass the radius of closest approach to those bodies, and trajectories that remain far from the radius of closest approach experience negligible deflection. This can be seen by computing the (dimensionless) ratio of the impact parameter to the distance $\Delta x_\mathrm{initial}$ or $\Delta x_\mathrm{end}$ between the spacecraft and the body at the respective initial or final time, whichever is closer and is therefore subject to a greater gravitational effect. These ratios are provided in the last column of Table \ref{table:OnePartMisses} for the bodies that do not pass the radius of closest approach $r_\mathrm{c}$. For the Earth, the small ratio of $b/\Delta x$ indicates that the body starts far away from $r_\mathrm{c}$ before moving further away, so one may indeed expect a small deflection. The case of Alpha Centauri A+B is milder (with a ratio $b/\Delta x$ closer to unity), but still exhibits a difference of an order of magnitude between the numerical result and theoretical estimate.  However, the distance between Alpha Centauri A+B and the target (Proxima Centauri b) is rather large ($\sim 8850~\mathrm{AU}$), so that one might expect to get a significant deflection only when the ratio $b/\Delta x$ is very close to unity.



The theoretical $\mathcal{O}(G^2)$ estimates provide a measure for the errors that arise from employing a first PM approximation to GR. In each individual case, the errors are many orders of magnitude smaller than the computed miss distances and the $\mathcal{O}(G)$ estimates. This indicates that the PM approximation includes the dominant gravitational contributions to the spacecraft trajectory. When comparing the results for different bodies, we see that the $\mathcal{O}(G^2)$ estimate for the Sun is comparable to the computed miss distance for Proxima Centauri, and exceeds that of the Moon and Mars. However, we note that $10^{-7}~\mathrm{AU}\approx 15~\mathrm{km}$, so the inclusion of higher order effects in GR is only necessary for accuracies on the order of a few tens of kilometers or smaller.

\begin{table*}[htbp]
\centering
{\renewcommand{\arraystretch}{1.4}%
 \begin{tabular}{|l @{\extracolsep{\fill}}|c|c|c|c|c|} 
 \hline
 Body & \texttt{PoMiN} & Newtonian & Theoretical $\mathcal{O}(G)$ & Theoretical $\mathcal{O}(G^2)$ & $b/\Delta x_\mathrm{initial}~(b/\Delta x_\mathrm{end})$ \\
 \hline\hline
 Sun
 & $1.393122 \times 10^{-1}$ & $1.354194 \times 10^{-1}$
 & $1.653598 \times 10^{-1}$ & $2.243105 \times 10^{-7}$
 & $8.926179\times 10^{-1}$ \\
 \hline
 Jupiter 
 & $1.524705 \times 10^{-5}$ & $1.513906 \times 10^{-5}$
 & $3.470577 \times 10^{-5}$ & $9.880820 \times 10^{-15}$
 & 
 ---\\
 \hline
 Earth
 & $1.985084 \times 10^{-6}$ & $1.988337 \times 10^{-6}$
 & $1.226868 \times 10^{-2}$ & $1.234768 \times 10^{-9}$
 & $3.370195\times 10^{-4}$ \\
\hline
 Alpha Centauri A+B ~~
 & $1.002445 \times 10^{-6}$ & $1.193681 \times 10^{-6}$
 & $2.675328 \times 10^{-5}$ & $5.871427 \times 10^{-15}$
 & $\left(8.122028\times 10^{-1}\right)$ \\
 \hline
 Proxima Centauri
 & $4.00097\times 10^{-7}$ & $4.583294\times 10^{-7}$
 & $3.011034 \times 10^{-8}$ & $7.614718\times 10^{-15}$
 & $\left(1.000006\right)$ \\
 \hline
 Moon
 & $2.483183 \times 10^{-8}$ & $2.487240 \times 10^{-8}$
 & $2.438216 \times 10^{-8}$ & $4.876790 \times 10^{-17}$ 
 & 
 ---\\
 \hline
 Mars
 & $1.641622 \times 10^{-8}$ & $1.611182 \times 10^{-8}$
 & $2.402962 \times 10^{-8}$ & $4.736781 \times 10^{-21}$ 
 & 
 ---\\
 \hline
 \end{tabular}
 }
 \caption{Miss distance due to an individual gravitating body, in AU, given in decreasing order of influence.  The first two columns list the displacement of the endpoint due to the gravity of an individual body, in AU, at a time $t_\mathrm{cl}$. The theoretical estimates are rough upper limits obtained from the scattering problem in the Schwarzschild spacetime.  The last column computes the (dimensionless) ratio of the impact parameter to the distance $\Delta x_\mathrm{initial}$ (or $\Delta x_\mathrm{end}$, in parentheses) between the spacecraft and the body at the respective initial or final time, whichever is closer, for the bodies that do not pass the radius of closest approach $r_\mathrm{c}$. Smaller values of the ratio generally indicate larger discrepancies between the computed values and $\mathcal{O}(G)$ estimate.}
 \label{table:OnePartMisses}
\end{table*}

\subsection{Six-body Fine-tuning}
\label{sect:fine-tuning}
The second numerical simulation involves a six-body calculation that implements the fine-tuning methods described in section \ref{sec:FineTuning}, which serves as a proof of concept for these methods. Given the results in Table \ref{table:OnePartMisses}, we perform a simulation with the five bodies that have the largest endpoint displacements, plus the spacecraft as a test particle. In particular, we consider a system consisting of the spacecraft, Sun, Jupiter, Earth, Alpha Centauri A+B, and Proxima Centauri, with the initial data supplied in section \ref{sec:InitialData}.  We then apply the root-finding methods of section \ref{sec:FineTuning} to the initial velocity of the spacecraft to find the precise value of the initial velocity that the spacecraft must have in order to pass through the target point at the time $t_\mathrm{cl}$. The fine-tuned initial data obtained in this manner are presented in Table \ref{table:FTIDComponents}.

Given fine-tuned initial data for the spacecraft, we rerun the simulation and compute the distance between the final position of the spacecraft and the target. The results are presented in the first two rows of Table \ref{table:FineTuning}. We obtain close to femtometer-scale miss distances, roughly the size of atomic nuclei. We stress that in each case, these miss distances neglect some physical effects: in the Newtonian case, relativistic effects are neglected, and in the relativistic (PM) case, higher order $\mathcal{O}(G^2)$ effects of GR (which can contribute errors as large as $\sim30~\mathrm{km}$ due to the Sun) are neglected. The small errors in Table \ref{table:FineTuning} are therefore indicative only of the achievable precision of the fine-tuning method within a given model.

We also rerun the relativistic simulation with fine-tuned Newtonian initial data to determine the size of the error that one might incur if one were to model the trajectory in a purely Newtonian manner. The resulting miss distance is presented in the third line of Table \ref{table:FineTuning}. We find that if one obtains a solution to the targeting problem with purely Newtonian considerations, the resulting spacecraft trajectory will miss the target by a distance of about 690,000 km, or about 21\% greater than the closest approach of the Voyager 2 probe to Jupiter \cite{Stone:1979voyager} and nearly twice the distance between the Earth and the Moon \cite{Bender:1973lunar}. If one seeks a targeting accuracy smaller than this, then one must account for relativistic effects.

\begin{table}[htbp]
\centering
{\renewcommand{\arraystretch}{1.4}%
 \begin{tabular}{|c|c|} 
 \hline
 Comp. & Value \\
 \hline\hline
 $x_{0,\mathrm{SC}}$
 & {\small $-2.235888445902370 \times 10^{7}~G M_\odot/c^2$} \\
 \hline
 $y_{0,\mathrm{SC}}$
 & {\small $8.680664864663420 \times 10^{7}~G M_\odot/c^2$} \\
 \hline
 $z_{0,\mathrm{SC}}$
 & {\small $2.988257878660500 \times 10^{7}~G M_\odot/c^2$} \\ 
 \hline\hline
 $v_{x,\mathrm{Rel.}}$
 & {\small $-7.29280133630346695175238301027084351 \times 10^{-2}~c$} \\
 \hline
 $v_{y,\mathrm{Rel.}}$
 & {\small $-5.57596267171519947482140178590898884 \times 10^{-2}~c$} \\
 \hline
 $v_{z,\mathrm{Rel.}}$
 & {\small $-1.77686212323486749509362995970259857 \times 10^{-1}~c$} \\
 \hline\hline
 $v_{x,\mathrm{Newt.}}$
 & {\small $-7.2928011848008478283629726326358021 \times 10^{-2}~c$} \\
 \hline
 $v_{y,\mathrm{Newt.}}$
 & {\small $-5.57596297969013443215818900874537475 \times 10^{-2}~c$} \\
 \hline
 $v_{z,\mathrm{Newt.}}$
 & {\small $-1.77686212073603716257005595684594549 \times 10^{-1}~c$} \\
 \hline
 \end{tabular}
 }
 \caption{Fine-tuned initial data for the spacecraft. The first three rows are coordinate values for the initial position $(x_{0,\mathrm{SC}},y_{0,\mathrm{SC}},z_{0,\mathrm{SC}})$, and the remaining rows are the fine-tuned velocity components for the relativistic case $(v_{x,\mathrm{Rel.}},v_{y,\mathrm{Rel.}},v_{z,\mathrm{Rel.}})$ and the Newtonian case $(v_{x,\mathrm{Newt.}},v_{y,\mathrm{Newt.}},v_{z,\mathrm{Newt.}})$.}
 \label{table:FTIDComponents}
\end{table}

\begin{table}[htbp]
\centering
{\renewcommand{\arraystretch}{1.4}%
 \begin{tabular}{|l @{\extracolsep{\fill}}|c|c|} 
 \hline
 Case & Miss in $\mathrm{AU}$ & Miss in $\mathrm{km}$ \\
 \hline\hline
 Relativistic
 & $7.1831 \times 10^{-26}$ & $1.07457 \times 10^{-17}~$\\
 \hline
 Newtonian
 & $3.30856 \times 10^{-26}$ & $4.9495 \times 10^{-18}$\\
 \hline
 Rel. w/ Newt. data~~ 
 & $4.6136 \times 10^{-3}$ & $690,180$\\
 \hline
 \end{tabular}
 }
 \caption{Miss distances with fine-tuned initial data. In both the purely relativistic case and the purely Newtonian case, we obtain machine precision values (using the extended precision floating point type supplied in $\texttt{DoubleFloats.jl}$ \cite{DoubleFloats2018}). When we use the fine-tuned Newtonian data with the relativistic Hamiltonian, we obtain a miss distance of $690,180~\mathrm{km}$.}
 \label{table:FineTuning}
\end{table}

\subsection{Endpoint Dispersion}
The last set of simulations involve another six-body calculation in which we investigate the combined effects of gravity and the random variation in the initial velocity of the spacecraft.  As in section \ref{sect:fine-tuning}, the six bodies include the spacecraft, the Sun, the Earth, Jupiter, Proxima Centauri, and Alpha Centauri A+B.

A baseline velocity vector is used for the spacecraft and taken from the fine-tuned velocity described in section \ref{sect:fine-tuning} and given in Table \ref{table:FTIDComponents}.  That baseline velocity is the one that results in a direct hit on Proxima Centauri b in the presence of the given gravitating bodies.

From that baseline velocity, perturbations are introduced using the method described in section \ref{sect:random_velocity}.  We run three sets of simulations: first, with an initial variation of $10^{-5}$ degrees ($30.6$ mas) in the initial velocity angle (corresponding to a miss distance of Proxima Centauri b of up to about $7,000,000$ km based on geometry alone); second, with an initial variation of $10^{-6}$ degrees ($3.6$ mas) in the initial velocity angle (corresponding to a miss distance of Proxima Centauri b of up to about $700,000$ km based on geometry alone); and third, with an initial variation of $10^{-7}$ degrees ($0.36$ mas) in the initial velocity angle (corresponding to a miss distance of Proxima Centauri b of up to about $70,000$ km based on geometry alone).  For each case, 200 initial velocities are generated and the miss distances are computed using \texttt{PoMiN}.

Table \ref{table:dispersion} contains the results.  Each reported miss distance is the mean after 200 trials.

The geometric miss distances are computed using the small angle approximation alone (expected miss distance = $d \cdot \theta$, where $d$ is the distance to the target and $\theta$ is the variation in launch angle).  The actual miss distances are computed using the first PM approximation to GR implemented in \texttt{PoMiN}.

The table also reports the difference between the actual and geometric miss distances.  These differences are small, on the order of a few meters for the $10^{-7}$ degree case up to 1.67 km for the $10^{-5}$ degree case.  This indicates that, even when the targeting problem is properly solved, the miss distance is dominated by the error in the initial angle produced during the launch and boost of the spacecraft.  Furthermore, the differences are negative, indicating that the actual misses are smaller than the geometric estimates of the misses by a fraction of a percent.  We attribute this slight ``focusing" effect to the gravitational influence of Proxima Centauri which attracts the spacecraft toward its center of mass at the end of the trajectory.

\begin{table*}[htbp]
\centering
{\renewcommand{\arraystretch}{1.4}%
 \begin{tabular}{|l|c|c|c|} 
 \hline
 Angular Variation $\theta$ (degrees) & Geometric miss distance & Actual miss distance  & Difference (actual - geometric) \\
 \hline\hline
 $10^{-5}$ 
 & $4.7679943\times 10^6~\mathrm{km}$ & $4.7679960 \times 10^6~\mathrm{km}$ & $-1676.1~\mathrm{m}$\\
 \hline
 $10^{-6}$ 
 & $4.6505606 \times 10^5~\mathrm{km}$ & $4.6505615 \times 10^5~\mathrm{km}$ & $-86.458~\mathrm{m}$\\
 \hline
 $10^{-7}$ 
 & $4.7529148 \times 10^4~\mathrm{km}$ & $4.7529155\times10^4~\mathrm{km}$ & $-6.9189~\mathrm{m}$\\
 \hline
 \end{tabular}
 }
 \caption{Miss distance as a result of angular variation in initial spacecraft launch velocity.  Each miss distance is the mean of 200 random trials.  Geometric miss distance is calculated geometrically (using the small angle approximation) based on the angular variation alone.  Actual miss distance is computed using the first PM approximation to GR.  From this, one can conclude that miss distances are dominated by the angular variation in the initial trajectory.}
 \label{table:dispersion}
\end{table*}

%
%
%

\section{Conclusions and Future Work}\label{sec:conclusions}

In this article, we established the gravitational effects on the trajectories of relativistic spacecraft traveling to interstellar destinations.  We performed numerical simulations to this end.

This study marked the first use of the re-implementation of \texttt{PoMiN} in the Julia language, which can be found in \cite{Feng:2018_17246109}.  Among other advantages, Julia provides the ability to perform algorithmic differentiation of numerical ODE solutions with respect to initial data.  This enabled us to develop an automated way -- using root-finding of a numerical miss function -- to select an ideal initial trajectory vector for hitting a target like Proxima Centauri b.  Combined with high-precision floating point types, this root-finding method can achieve a target precision of a femtometer over a distance of 4.25 light years, within a given model.

When considering the gravitational influence of individual bodies, including the Sun, Earth, Moon, Jupiter, Mars, Proxima Centauri, and Alpha Centauri A+B (see Table \ref{table:OnePartMisses}), we found that the Sun had the largest influence on the spacecraft's trajectory.

Considering the bodies with the largest gravitational effects, we found that neglecting relativistic effects (and only including Newtonian modeling) gave us a miss distance of about 690,000 kilometers when targeting Proxima Centauri b (see Table \ref{table:FineTuning}).  For context, this miss distance is about 21\% larger than the closest approach of Voyager 2 to Jupiter.

We also evaluated the errors arising from the leading post-Minkowskian (PM) approximation to general relativity (GR) for scattering off of individual bodies (see Table 1), and found that in each individual case the estimated contribution from higher order effects in GR is orders of magnitude smaller than that obtained in the first PM approximation.  We therefore conclude that the first PM approximation includes the dominant gravitational contributions to the spacecraft trajectory.  We note that the higher order contribution from full GR in the case of the Sun is comparable to the first PM effect from the gravitational influence of Proxima Centauri.  However, this effect amounts to an error on the order of tens of kilometers.  Unless a target precision better than tens of kilometers is required, the first PM approximation, in the form of the LSB Hamiltonian in Eqs. (\ref{LSBHamiltonian}--\ref{LSBHamiltonianH3}), is a suitable model to use in place of GR.  

Lastly, we investigated the effect of variation in the angle of the initial velocity that could arise from errors during the spacecraft launch and boost phase.  We found that the target miss distances are dominated by the variation in the initial angle (confirming the results of \cite{Jackson:2017disp}).  Gravity reduces the miss distances by a very small amount, due to a gravitational ``focusing" effect that is likely due to the gravitational attraction of Proxima Centauri at the trajectory's endpoint.

An effort is currently underway to implement external gravitational fields in the relativistic Hamiltonian in \texttt{PoMiN}, which would enable one to include the effects of the Milky Way potential on interstellar spacecraft trajectories.  And, in the case where a precision better than tens of kilometers is needed for hitting the target, a more complete general-relativistic description of the Sun's gravitational field could be added as a background field alongside the first PM gravitational dynamics.


%
%


\section*{Acknowledgments}
We thank Al Jackson and Richard Matzner for bringing relativistic interstellar spacecraft to our attention as a possible area of application of our \texttt{PoMiN} methodology. J. C. F. is supported by the European Union and Czech Ministry of 
Education, Youth and Sports through the FORTE project No. 
CZ.02.01.01/00/22 008/0004632.


\bibliography{ref}


%
%

%
%


%
%

%
%

\end{document}